\documentclass[aps,pra,reprint,superscriptaddress]{revtex4-2}

\usepackage{bm}
\usepackage{color}
\usepackage[dvipdfmx]{graphicx}

\newcommand{\ahat}{\hat{a}}
\newcommand{\ahattwo}{\hat{a}^2}
\newcommand{\adag}{\hat{a}^\dagger}
\newcommand{\adagtwo}{\hat{a}^{\dagger 2}}
\newcommand{\ave}[1]{\langle #1 \rangle}

\newcommand{\bra}[1]{\langle #1 |}
\newcommand{\Cb}{C_{\text{b}}}
\newcommand{\Cs}{C_{\text{s}}}
\newcommand{\fsp}{f_{\text{sp}}}
\newcommand{\fsq}{f_{\text{sq}}}
\newcommand{\Hb}{\hat{H}_{\text{b}}}
\newcommand{\Hbi}{\hat{H}_{\text{b}, i}}
\newcommand{\Hbmf}{\hat{H}_{\text{b}}^{\text{mf}}}
\newcommand{\Hbtwo}{\hat{H}_{\text{b}}^{(2)}}
\newcommand{\Hs}{\hat{H}_{\text{s}}}
\newcommand{\Hsi}{\hat{H}_{\text{s}, i}}
\newcommand{\Hsmf}{\hat{H}_{\text{s}}^{\text{mf}}}
\newcommand{\HsN}{\hat{H}_{\text{s}}^{(N)}}
\newcommand{\Hstwo}{\hat{H}_{\text{s}}^{(2)}}
\newcommand{\Hstwop}{\hat{H}'^{(2)}_{\text{s}}}
\newcommand{\ket}[1]{|#1\rangle}
\newcommand{\psib}{\psi_{\text{b}}}
\newcommand{\psis}{\psi_{\text{s}}}
\newcommand{\sigmaikx}{\hat{\sigma}_{i,k}^x}
\newcommand{\sigmaiky}{\hat{\sigma}_{i,k}^y}
\newcommand{\sigmaikz}{\hat{\sigma}_{i,k}^z}
\newcommand{\sigmajlz}{\hat{\sigma}_{j,l}^z}
\newcommand{\ssq}{\hat{\bm{s}}^2}
\newcommand{\Sx}{\hat{S}^x}
\newcommand{\Sy}{\hat{S}^y}
\newcommand{\Sz}{\hat{S}^z}
\newcommand{\sx}{\hat{s}^x}
\newcommand{\sy}{\hat{s}^y}
\newcommand{\sz}{\hat{s}^z}
\newcommand{\tJ}{\tilde{J}}
\newcommand{\tp}{\tilde{p}}
\newcommand{\tDelta}{\tilde{\Delta}}
\newcommand{\tepsilon}{\tilde{\epsilon}}

\begin{document}

\title{Effective spin models of Kerr-nonlinear parametric oscillators 
for quantum annealing}


\author{Ryoji Miyazaki}
\affiliation{Secure System Platform Research Laboratories, NEC Corporation, Kawasaki 211-8666, Japan}
\affiliation{NEC-AIST Quantum Technology Cooperative Research Laboratory,
National Institute of Advanced Industrial Science and Technology (AIST), Tsukuba, Ibaraki 305-8568, Japan}


\date{\today}

\begin{abstract}
A method of quantum annealing (QA)
using Kerr-nonlinear parametric oscillators (KPOs) 
was proposed.
This method is described by bosonic operators 
and has different characteristics 
from QA based on the transverse-field Ising model.
As the first step 
to describe this method in the conventional framework of QA, 
we propose effective spin models of KPOs.
The spin models are obtained via a variant of the Holstein-Primakoff transformation 
and are described by spin-$s$ operators.
The terms for detuning, coherent driving, parametric driving, 
and the Kerr effect 
are mapped to 
the transverse field, longitudinal field, 
nonlinear terms for the z- and x-components of spins, 
respectively.
By analyzing the spin models 
corresponding to KPOs in several settings, 
we demonstrate that 
the present spin models for a rather large $s$ 
and tuned parameters qualitatively reproduce behavior of KPOs.
\end{abstract}

\maketitle

\section{Introduction}

Quantum annealing (QA) is a heuristic to solve combinatorial optimization problems
by embedding the problems 
in systems with quantum fluctuations~\cite{T.Kadowaki1998, A.Das2008, T.Albash2018Jan, P.Hauke2020}.
In the ideal scenario for QA, 
the system follows the instantaneous ground state 
during the gradual decrease of quantum fluctuations
and finally reaches the ground state 
that corresponds to the solution of the problem.
It is natural to utilize the Ising model, 
which represents the problems~\cite{A.Lucas2014}, 
with simple extra terms for quantum fluctuations.
The transverse-field Ising model (TFIM) is such a model 
and has been typically used 
in studies of QA~\cite{A.Das2008, T.Albash2018Jan, P.Hauke2020}.
Such studies have been also inspired by the advent of quantum annealers 
provided by D-Wave Systems Inc.~\cite{M.Johnson2011}, 
which are well described by the TFIM.
Variants of the model including other terms for quantum fluctuations, 
e.g., non-stoquastic catalysts~\cite{Y.Seki2012, B.Seoane2012, Y.Seki2015, H.Nishimori2017, L.Hormozi2017, Y.Susa2017, A.Ozguler2018, T.Albash2019Apr2, K.Takada2020}, 
have been also investigated.
Thus QA has been studied primarily 
in terms of the TFIM and its variants.

QA, however, does not have to be described by spin systems.
Indeed, a method related to QA was proposed 
using 
Kerr-nonlinear parametric osicllators (KPOs)~\cite{H.Goto2016Feb, H.Goto2019Mar}.
KPOs are represented by bosonic operators 
and have continuous degrees of freedom, 
but final states can be associated with the Ising model~\cite{H.Goto2016Feb, H.Goto2019Mar}.
The pump amplitude for the parametric driving is gradually increased 
in this method.
This adiabatic process for a KPO generates the cat state, 
i.e., superposition of two coherent states
with opposite phases~\cite{H.Goto2016Feb, H.Goto2019Mar, S.Puri2017Apr} 
in which 
we can encode the up and down states of an Ising spin. 
The underlying mechanism to obtain Ising spins is interpreted 
as bifurcation~\cite{H.Goto2016Feb, H.Goto2019Mar}.
The ground state of the Ising model can be obtained by 
applying this method to networks of interacting KPOs. 
We consider this method in terms of a kind of QA
and simply call it QA with KPOs hereafter.

QA with KPOs has been intensively studied~\cite{H.Goto2016Feb, H.Goto2019Mar, S.Puri2017Jun, T.Kanao2021, P.Zhao2018, H.Goto2018, Y.Zhang2017, H.Goto2020, S.Nigg2017, M.Dykman2018, M.Kewming2020, T.Onodera2020}.
The efficiency of this method was demonstrated 
with simulations of small systems~\cite{H.Goto2016Feb, H.Goto2019Mar, S.Puri2017Jun}.
Subsequent studies pointed out the possibility 
of applications to 
the Lechner-Hauke-Zoller scheme~\cite{W.Lechner2015} 
with four- or three-body interactions~\cite{S.Puri2017Jun, T.Kanao2021, P.Zhao2018}, 
Boltzmann sampling~\cite{H.Goto2018}, 
and QA using excited states~\cite{Y.Zhang2017, H.Goto2020}.
The bifurcation-based approach to QA
has also been studied using a spin model~\cite{K.Takahashi2022}.
A KPO is not just a theoretical model 
and is realized in superconducting circuits 
with the Josephson junctions~\cite{Z.Wang2019, A.Grimm2020, T.Yamaji2022}.
The physical implementation of QA with KPOs 
is also expected~\cite{S.Nigg2017, S.Puri2017Jun, P.Zhao2018, T.Onodera2020}.
Qubits generated by the adiabatic process for KPOs 
can be utilized for the gate-based quantum computation~\cite{H.Goto2016May, S.Puri2017Apr, A.Grimm2020, S.Puri2020, A.Darmawan2021, Q.Xu2022, H.Putterman2022, T.Kanao2021Augarxiv}.

One of the characteristics of QA with KPOs 
is being described by bosonic operators.
This point highlights a difference from conventional QA based on the TFIM
but also makes it difficult to understand this method with the concepts
developed in the study of the conventional QA.
It is a natural question 
how this method can be described in terms of the conventional QA.
The first step to answer this question is 
to construct effective spin models for KPOs.
The models should be as simple as the TFIM.
Such spin models enable us to discuss the two methods at the same stage, 
where the systems are represented by spins. 
The spin representation would more clearly show the role of each parameter of KPOs 
in QA.
In this regard, 
identifying the correspondence to spin models is similar in purpose 
to the gate synthesis problem discussed in gate-based quantum computation~\cite{A.Magann2021, B.Ozguler2022arxiv}.
Note that 
a spin model was proposed for the bifurcation-based QA 
as mentioned above~\cite{K.Takahashi2022}, 
but this model was not derived from KPOs.
An effective spin model for KPOs could clarify the relation 
between the previously proposed spin model and KPOs.

In this paper, 
we propose effective spin models for KPOs.
We aim to find simple spin models 
that qualitatively reproduce behavior of KPOs 
and clarify roles of parameters of KPOs in QA.
The next section gives a brief review of QA with KPOs.
In Sec.~\ref{sec:transform}, 
the spin models are derived 
via the Holstein-Primakoff (HP) transformation~\cite{T.Holstein1940, A.Auerbach1998}.
In Sec.~\ref{sec:comparison}, 
we compare physical quantities such as photon number 
computed for the bosonic models of KPOs 
and their spin counterparts computed for the corresponding spin models.
We present conclusion 
in Sec.~\ref{sec:summary}

\section{Quantum annealing with Kerr-nonlinear parametric oscillators}

KPOs are typically investigated in the frame rotating at half the frequency of the parametric driving.
A KPO in the frame under the rotating-wave approximation 
is governed by~\cite{H.Goto2016Feb, H.Goto2019Mar}
\begin{equation}
\Hb 
= \Delta \adag \ahat + \frac{K}{2} \adagtwo \ahattwo 
-\frac{p}{2} (\adagtwo + \ahattwo)
- \epsilon (\adag + \ahat), 
\label{eq:H_KPO}
\end{equation}
where $\adag$ and $\ahat$ are bosonic creation and annihilation operators, respectively. 
The subscript b emphasizes that
this is a Hamiltonian for bosons.
This is a model, for example, 
of superconducting circuits 
with the Josephson junctions~\cite{Z.Wang2019, A.Grimm2020, T.Yamaji2022}.
The parametric driving with the pump amplitude $p$ 
is controlled by the modulated magnetic flux through SQUIDs.
The Kerr nonlinearlity $K$ stems from the nonlinearlity of the Josephson junctions.
The detuning $\Delta$ represents the difference of 
the oscillator frequency from half the pump frequency.
The coherent driving is applied with amplitude $\epsilon$ to generate the bias 
in the system.
However, 
the investigation in this paper is not restricted to specific implementations 
to superconducting circuits.
Note that $\hbar = 1$ throughout the paper.

QA with KPOs is carried out 
by varying the pump amplitude~\cite{H.Goto2016Feb}.
We prepare the ground state of the system for $p = 0$ and $\Delta > 0$
and gradually increase $p$.
If the energy gap between the ground state and excited states does not close 
in the annealing passage, 
the system will track the instantaneous ground state for temporal $p$~\cite{T.Albash2018Jan}.
Consequently, 
we will obtain the ground state of the KPO for large $p$.
Let us assume $\epsilon = 0$ for simplicity. 
The initial state is the vacuum state.
For large $p$, where $\Delta$ can be ignored, 
the Hamiltonian is written as
\begin{equation}
\Hb
= \frac{K}{2} \left( \adagtwo - \alpha_0^2 \right) \left( \ahattwo - \alpha_0^2 \right), 
\end{equation}
where $\alpha_0 = \sqrt{p/K}$.
The ground state obtained by QA starting from the vacuum state 
is the so-called even cat state, 
$(\ket{\alpha_0} +\ket{-\alpha_0})/\sqrt{2}$, 
where a coherent state $\ket{\alpha}$ is defined as 
$\ahat \ket{\alpha} = \alpha \ket{\alpha}$~\cite{D.Walls2008}.
Both $\ket{\alpha_0}$ and $\ket{-\alpha_0}$ are eigenstates of the Hamiltonian
with the same eigenvalue.
The final state is their superposition 
and has the same parity as the initial state, 
namely the even parity, 
since the Hamiltonian commutes the parity operator 
that leads to $\ahat \to -\ahat$. 
If we also gradually increase the coherent driving, 
the ground state for large $p$ is not the cat state
but a state localized in one of the double wells 
in the meta-potential for the KPO.
If the absolute value of $\epsilon$ is small compared to $p$, 
the state is close to $\ket{\alpha_0}$ for $\epsilon>0$ 
and $\ket{-\alpha_0}$ for $\epsilon<0$~\cite{S.Puri2017Jun}.
This relation is reminescent of an Ising spin 
in a magnetic field~\cite{H.Nishimori2011}.
In addition, 
$\ket{\alpha_0}$ and $\ket{-\alpha_0}$ for large $\alpha_0$ 
are approximately orthogonal, 
since 
$\langle \alpha_0 | -\alpha_0 \rangle = e^{-2\alpha_0^2}$~\cite{D.Walls2008}.
Therefore, 
we can encode an Ising spin $\{ \uparrow, \downarrow \}$ 
in the states $\{ \ket{\alpha_0},  \ket{-\alpha_0} \}$
and interpret the coherent driving as the magnetic field applied to the spin.

The Ising model is artificially realized 
by networks of interacting KPOs~\cite{H.Goto2016Feb},  
\begin{equation}
\Hb^{(N)} 
= \sum_{i=1}^N \Hbi
-\xi_0 \sum_{i,j}J_{ij} \adag_i \ahat_j, 
\label{eq:H_KPOs} 
\end{equation}
\begin{equation}
\Hbi
= \Delta \adag_i \ahat_i + \frac{K}{2} \adagtwo_i \ahattwo_i
-\frac{p}{2} (\adagtwo_i + \ahattwo_i)
- \epsilon_i (\adag_i + \ahat_i), 
\label{eq:H_KPO_i}
\end{equation}
where $J_{ij}$ will correspond to the coupling constant of Ising spins $i$ and $j$.
The parameter $\xi_0$ is introduced 
to satisfy that
the vacuum state is the ground state of the initial Hamiltonian 
for $p = 0$ and $\epsilon = 0$~\cite{H.Goto2016Feb, H.Goto2019Mar}.
For sufficiently large $p$, 
where $\Delta$ is ignored, 
each KPO in the ground state is approximately described 
by one of the two coherent states $\ket{\sigma \alpha_0}$ 
with $\sigma = \pm 1$ representating an Ising spin. 
Hence, the ground state of the whole system can be expressed as 
$\ket{\bm{\sigma}} = \ket{\sigma_1 \alpha_0}\cdots \ket{\sigma_N \alpha_0}$. 
The expectation value of the energy for $\ket{\bm{\sigma}}$ is 
\begin{equation}
\bra{\bm{\sigma}} \Hb^{(N)} \ket{\bm{\sigma}}
= -2 \alpha_0 \sum_{i=1}^N \epsilon_i \sigma_i
-\xi_0 \alpha_0^2 \sum_{i,j} J_{ij} \sigma_i \sigma_j, 
\end{equation}
where constants are dropped.
This relation maps the ground state of the KPOs to that of the Ising model.
We can obtain the ground state of the KPOs with the gradual increase of $p$. 
Thus, the ground state of the Ising model is found  
with this scheme.
This mapping to Ising spins requires the ground state 
to be composed of the coherent states with the same amplitude.
This is guaranteed for large $p$, 
but the system for not large $p$ is not necessarily 
in such a state.
We need to assume more complicated states to describe the system 
in the intermediate range of $p$.

\section{Transformation}
\label{sec:transform}

We utilize the Holstein-Primakoff (HP) transformation~\cite{T.Holstein1940, A.Auerbach1998} 
to obtain spin models for KPOs.
The HP transformation is defined by
\begin{eqnarray}
\hat{S}^+ 
&&= \sqrt{2S - \adag \ahat} \ahat, 
\\
\hat{S}^- 
&&= \adag \sqrt{2S - \adag \ahat}, 
\\
\Sz 
&&= S - \adag \ahat, 
\label{eq:Sz_HP}
\end{eqnarray}
where $\hat{S}^\pm = \Sx \pm i \Sy$, 
and $\Sx$, $\Sy$, and $\Sz$ are the x-, y-, and z-components 
of the spin-$S$ operator.
$S$ is an integer or half-integer.
The spin operators obey 
$[\hat{S}^\alpha, \hat{S}^\beta] = i \epsilon^{\alpha\beta\gamma} \hat{S}^\gamma$, 
where $\alpha$, $\beta$, $\gamma$ run over $x$, $y$, $z$, 
and $\epsilon^{\alpha\beta\gamma}$ is the totally antisymmetric tensor.
$[\ahat, \adag] = 1$ also holds.
We can map spin systems to bosonic ones
with the above relation, 
but the Fock states for the bosonic ones with photon number larger than $2S$ 
are unphysical states for the original spin systems.
Therefore, we only consider the subspace 
spanned by the Fock states with photon number equal to or smaller than $2S$, 
while the other states are eliminated by a projector.
In typical treatment~\cite{T.Holstein1940, A.Auerbach1998}, 
the square root for $S^+$ or $S^-$ is expanded in powers of $1/S$,
\begin{equation}
\hat{S}^+ 
= \sqrt{2S} 
\left[ 1 - \frac{\adag \ahat}{4S} 
- \frac{(\adag \ahat)^2}{32S^2} 
- \cdots \right] \ahat.
\label{eq:S+_HP}
\end{equation} 
This can be interpreted as the expansion in powers of $\adag \ahat/2S$.
Other similar approaches~\cite{A.Klein1991}
such as the Schwinger bosons 
and the Dyson-Maleev transformation 
can be applied, 
but the former requires multiple boson species to represent a spin, 
and the latter breaks Hermeticity.
We therefore adopt the HP transformation.
Note that 
a more sophisticated version of the HP transformation 
has recently been proposed~\cite{M.Vogl2020, J.Konig2021}, 
but we concentrate in this paper on examining the approach 
based on the original one.

We use the HP transformation in the opposite direction
to transform KPOs to spin models.
In addition, we apply the gauge transformation, 
$(\Sx, \Sy, \Sz) \to (\sz, -\sy, \sx)$, 
which preserves the spin commutation relation.
Here, spin operators for the latter gauge are represented with lowercase letters  
to distinguish the gauges.
For example, the vacuum state, 
which is the initial state in QA with KPOs, 
is mapped to the eigenstate $\ket{m=s, s}_x$ of $\sx$ for eigenvalue $s$.
We also slightly modify the expansion 
and expand the square root in powers of $\adag \ahat/2s$ at $\alpha^2/2s$, 
where $\alpha^2$ will be set to some value.
Then 
$\hat{S}^+$ in Eq.~(\ref{eq:S+_HP}) is rewritten as
\begin{eqnarray}
\hat{s}^-
=&& \sz - i\sy
\nonumber \\
=&& \sqrt{2s - \adag \ahat} \ahat 
\nonumber \\
=&& \sqrt{2s} 
\Bigg[ \sqrt{1-\frac{\alpha^2}{2s}} - \frac{\adag \ahat - \alpha^2}{4s\sqrt{1 - \frac{\alpha^2}{2s}}} 
\nonumber \\
&&- \frac{(\adag \ahat - \alpha^2)^2}{32s^2 \left(1 - \frac{\alpha^2}{2s} \right)^{3/2}} 
- \cdots \Bigg] \ahat.
\label{eq:s-}
\end{eqnarray}
We also have
\begin{eqnarray}
\sx
&&= s - \adag \ahat, 
\label{eq:sx}
\\
\sy
&&=\frac{\sqrt{2s}}{i} 
\Bigg[ \sqrt{1 - \frac{\alpha^2}{2s}} \frac{\adag - \ahat}{2} \nonumber \\
&&- \frac{\adagtwo \ahat - \adag \ahat^2 - \alpha^2 (\adag - \ahat)}
{8s \sqrt{1 - \frac{\alpha^2}{2s}}} + \cdots \Bigg].
\label{eq:sy}
\\
\sz
&&= \sqrt{2s} 
\Bigg[ \sqrt{1 - \frac{\alpha^2}{2s}} \frac{\adag + \ahat}{2} \nonumber \\
&&- \frac{\adagtwo \ahat + \adag \ahat^2 - \alpha^2 (\adag + \ahat)}
{8s \sqrt{1 - \frac{\alpha^2}{2s}}} + \cdots \Bigg].
\label{eq:sz}
\end{eqnarray}
We use this relation 
to represent the terms in $\Hb$ [Eq.~(\ref{eq:H_KPO})]
with $\sx$ and $\sz$ as follows:
\begin{eqnarray}
\adag \ahat 
&&= s- \sx, 
\label{eq:adagahat}
\\
\adagtwo \ahattwo 
&&= (\sx)^2 -(2s-1)\sx + s(s-1), 
\label{eq:adagtwoahattwo}
\\
\frac{\adag + \ahat}{2} 
&&= \frac{1}{\sqrt{2s- \alpha^2}} \sz + \cdots, 
\label{eq:(adag+ahat)/2}
\\
\frac{\adagtwo + \ahattwo}{2} 
&&= - s + \sx - \frac{1}{2} + \frac{2}{2s- \alpha^2} (\sz)^2 + \cdots.
\label{eq:(adagtwo+ahattwo)/2}
\end{eqnarray}
Here, the value of $s$ is not specified, 
but $\alpha^2 < 2s$ is assumed.
The leading terms shown 
in Eqs.~(\ref{eq:(adag+ahat)/2}) and (\ref{eq:(adagtwo+ahattwo)/2}) 
are obtained from Eq.~(\ref{eq:sx}) and the first term in Eq.~(\ref{eq:sz}).
Their higher order terms obtained from the second term 
in Eq.~(\ref{eq:sz}) are shown in Appendix.

According to Eqs.~(\ref{eq:sx})--(\ref{eq:(adagtwo+ahattwo)/2}), 
$\Hb$ is transformed to 
\begin{eqnarray}
\Hs 
=&& - \frac{2p}{2s - \alpha^2} (\sz)^2
- \left[ \Delta + p + \left(s - \frac{1}{2} \right) K \right] \sx 
\nonumber \\
&&+ \frac{K}{2} (\sx)^2
-\frac{2\epsilon}{\sqrt{2s - \alpha^2}} \sz, 
\label{eq:H_spin}
\end{eqnarray}
where higher order terms and constants are dropped.
The subscript s emphasizes that
this is a Hamiltonian for spins.
Detuning $\Delta$ works as a transverse field, 
while the coherent driving $\epsilon$ turns to a longitudinal field.
The Kerr term generates the nonlinearlity of $\sx$.
Larger pump amplitude $p$ enhances fluctuations associated with $\sz$, 
which is what we mainly control in QA with KPOs 
and causes the bifurcation.
Since the nonlinear terms cannot be represented with spin-1/2 operators, 
$\Hs$ needs $s\ge 1$ to describe QA with the bifurcation mechanism.
$\Hs$  for $s=1$ without $(\sx)^2$ has the same form 
as a model for bifurcation-based QA~\cite{K.Takahashi2022}.
Thus $\Hs$ bridges a gap 
between KPOs and bifurcation-based QA with spin models 
by clarifying coefficients of the spin operators 
as a function of the parameters of KPOs.

The Hamiltonian of interacting KPOs [Eq.~(\ref{eq:H_KPOs})] is tranformed to
\begin{equation}
\HsN
= \sum_{i=1}^N \Hsi
-\frac{2\xi_0}{2s - \alpha^2} \sum_{i<j}J_{ij} 
\left( \sz_i \sz_j + \sy_i \sy_j \right), 
\end{equation}
where $\Hsi$ denotes the Hamiltonian for spin $i$ [Eq.~(\ref{eq:H_spin})].
Here, we have assumed $J_{ij} = J_{ji}$.
$\sz_i \sz_j$ might be more dominant than $\sy_i \sy_j$ 
in the coupling of spins $i$ and $j$, 
since flucutations for $\sz_i$ are enhanced with the pump in $\Hsi$.
In other words, 
it might be valid to approximate the spin Hamiltonian as
\begin{equation}
\hat{H}'^{(N)}_s
= \sum_{i=1}^N \Hsi
-\frac{2\xi_0}{2s - \alpha^2} \sum_{i<j}J_{ij} 
\sz_i \sz_j. 
\end{equation} 
We will examine the validity of this approximation later.
Note that 
if a bosonic system has $\adag_i \adag_j \ahat_i \ahat_j$, 
the corresponding spin system has the coupling of the x-components 
$\sx_i \sx_j$, 
but we do not discuss such a coupling in this paper.

We made two approximations, 
except to ignore $\sy_i \sy_j$, 
in the process of obtaining the spin systems.
One is the restriction of the Hilbert space.
The obtained spin systems take into account of only the Fock states 
whose photon number is equal to or smaller than $2s$, 
since states for larger photon number than $2s$ are unphysical ones 
for the spin systems.
If the state of a KPO includes the Fock states 
for a larger photon number than $2s$, 
the spin systems obtained with the transformation would not describe it well.
Larger $s$ could lead to a more accurate description of a KPO, 
but the spin model for large $s$ is far from the Ising model.
The other approximation is truncation of the expansion of square root.
The Hamiltonian in Eq.~(\ref{eq:H_spin}) is based 
only on the leading terms in the expansion. 
Including higer order terms could improve the accuracy 
of the transformation, 
while the spin model becomes more complicated.
The parameter $\alpha$ is introduced 
to reduce errors associated with this point.
To illustrate the motivation for the introduction of $\alpha$, 
we focus on the expansion in Eq.~(\ref{eq:s-}), 
which is represeted with a diagonal matrix in the occupation basis.
Some diagonal elements of $\adag \ahat$ are close to $\alpha^2$, 
and the corresponding elements of terms in the expansion are small.
Thus, 
when a state of a bosonic system can be well approximated 
by a superposition of only the Fock states 
with photon number close to $\alpha^2$, 
ignoring higher order terms is reasonable.
One state we investigate is the ground state of a KPO, 
in particular, 
without the coherent driving, 
which is the superposition of coherent states 
such as $(\ket{\alpha_0} +\ket{-\alpha_0})/\sqrt{2}$.
Both expectation and variance of $\adag \ahat$ 
for $\ket{\alpha_0}$ 
are $\alpha_0^2$~\cite{D.Walls2008}.
This fact means that
this state can be approximated by a superposition of only the Fock states 
with photon number close to $\alpha_0^2$
when $\alpha_0^2$ is small.
Therefore, we expect that
setting $\alpha^2$ to an approximate expectation value of $\adag \ahat$ 
for the ground state of a KPO reduces errors 
due to the truncation of expansion, 
especially when the amplitude of coherent states 
constituting the state is small. 

We next clarify the relationship between states for the bosonic and spin systems.
The transformation associates 
the Fock state $\ket{n}$ 
with an eigenstate $\ket{s-n, s}_x$ of $\sx$ for spin-$s$, 
where $\adag \ahat \ket{n} = n\ket{n}$ 
and $\sx\ket{s-n, s}_x = (s-n)\ket{s-n, s}_x$.
In particular, the vacuum state $\ket{0}$ corresponds to $\ket{s,s}_x$.
Let us focus on
the coherent state 
$\ket{\alpha_0} = \exp[\alpha_0 (\adag - \ahat)] \ket{0}$~\cite{D.Walls2008}.
For simplicity we consider the large $s$ limit, 
where Eq.~(\ref{eq:sy}) reduces to $\adag - \ahat = i \sqrt{2/s} \sy$.
We then have 
\begin{eqnarray}
\ket{\alpha_0}
&&= e^{i\theta_0\sy} \ket{s, s}_x
= \ket{\theta_0}, 
\label{eq:gs_KPO} 
\\
\theta_0
&&= \sqrt{\frac{2}{s}}\alpha_0.
\end{eqnarray}
The right hand side of Eq.~(\ref{eq:gs_KPO}) represents the spin coherent state 
$\ket{\theta_0}$, 
created by rotating the maximally polarized state $\ket{s, s}_x$ 
around the y axis with angle $\theta_0$~\cite{A.Auerbach1998}.
Thus the ground state of a KPO without the coherent driving, 
$(\ket{\alpha_0} + \ket{-\alpha_0})/\sqrt{2}$, 
is associated with the superposition of spin coherent states, 
$(\ket{\theta_0} + \ket{-\theta_0})/\sqrt{2}$, 
in the large $s$ limit.
Therefore in solving Ising problems with our spin models 
the states $\{ \uparrow, \downarrow \}$ of an Ising spin are encoded 
in the two spin coherent states $\{ \ket{\theta_0}, \ket{-\theta_0} \}$.

We should mention that 
the obtained spin model for $N$ KPOs can be interpreted 
as the model of spin-$1/2$ with $N (2s+1)$ spins, 
\begin{eqnarray}
\Hsi' 
=&& - \frac{2p}{2s - \alpha^2} 
\left( \sum_{k=1}^{2s+1} \sigmaikz \right)^2 \nonumber \\
&&- \left[ \Delta + p + \left(s - \frac{1}{2} \right) K \right] 
\sum_{k=1}^{2s+1} \sigmaikx \nonumber \\
&&+ \frac{K}{2} \left( \sum_{k=1}^{2s+1} \sigmaikx \right)^2
-\frac{2\epsilon}{\sqrt{2s - \alpha^2}} \sum_{k=1}^{2s+1} \sigmaikz, 
\end{eqnarray}
\begin{equation}
\Hs'^{(N)}
= \sum_{i=1}^N \Hsi'
-\frac{4\xi_0}{2s - \alpha^2} \sum_{i,j}J_{ij} \sum_{k, l=1}^{2s+1} \sigmaikz \sigmajlz, 
\end{equation}
where KPO $i$ is described by $2s+1$ spins, 
and $\sigmaikx$ and $\sigmaikz$ are the x- and z-components 
of $k$th spin-1/2 for KPO $i$.
If the initial state for QA with $\Hs'^{(N)}$ is set 
to an eigenstate of $\ssq$ for eigenvalue $s(s+1)$, 
where $\ssq = (\sum_{k=1}^{2s+1}\hat{\bm{\sigma}}_{i,k} )^2$ and 
$\hat{\bm{\sigma}}_{i,k} = (\sigmaikx, \sigmaiky, \sigmaikz)$, 
$\Hs'^{(N)}$ is effectively reduced to $\HsN$.
This is because $\Hs'^{(N)}$ commutes $\ssq$, 
and the state evolves in the subspace 
spanned by eigenstates of $\ssq$ for eigenvalue $s(s+1)$, 
where $\hat{\bm{s}}$ is equivalent to spin-$s$.
In this representation, 
Ising spin $i$ is encoded in the sign of the expectation value of $\sum_{k=1}^{2s+1} \sigmaikz$.

\section{Comparison of the models}
\label{sec:comparison}

\subsection{A single KPO}
\label{sec:1KPO}

We compare a KPO [Eq.~(\ref{eq:H_KPO})] 
and the corresponding spin models [Eq.~(\ref{eq:H_spin})], 
where higher order terms are ignored, 
for several values of $\alpha$ and $s$.
One of the values of $\alpha$ is deteremined by 
\begin{equation}
\alpha_c =
\left[ \tp-\tDelta + \frac{|\tepsilon|}{\sqrt{\left(\tp-\tDelta\right)}} \right]^{1/2}
\theta(\tp-\tDelta-\tepsilon).
\label{eq:alpha_c}
\end{equation}
Here, we have used $\tp = p/K$, $\tDelta = \Delta/K$, 
and $\tepsilon = \epsilon/K$.
$\theta(x) = 0$ if $x<0$, otherwise $\theta(x) = 1$.
$\alpha_c^2$ is an approximate expectation value of $\adag \ahat$ 
for the ground state of a KPO for small $|\tepsilon|$ 
based on the semiclassical analysis, 
where the system is assumed to be in a coherent state.
$\alpha_c$ is set to zero for $\tp - \tDelta - \tepsilon < 0$ 
to stabilize numerical calculations.
We focus on the ground states
and represent the expectation value of $\hat{X}$ for the ground state as $\ave{\hat{X}}$.
If $\hat{X}$ is a spin operator, 
the expectation value is calculated 
for the ground state of the spin model.
We investigate the models for $\alpha = 0$, $\alpha_c$, 
and equal to the exact $\sqrt{\ave{\adag \ahat}}$ of a KPO.
More specifically, 
we compare 
the photon number, quadrature amplitude, and the Wigner function, 
of the bosonic model and 
their spin counterparts.
For example, 
the spin counterpart of photon number $\ave{\adag \ahat}$ is $s - \ave{\sx}$, 
which we refer to as $\fsp$.
We are interested in a KPO for not so large $\tp$, 
where our transforamation works under the assumption $\alpha^2 < 2s$
with $\alpha = \alpha_c$. 
When we numerically analyze a KPO, 
we consider the Fock states whose photon number is smaller than 20.
We have confirmed that this truncation does not affect the result shown below.
Correspondingly, 
we investigate the spin models for $s \le 10$.
If $s = 10$, the spin model takes into account of only the Fock states 
whose photon number is equal to or smaller than 20. 
Hereafter, 
we use dimensionless parameters, i.e., $\tDelta$, $\tp$, and $\tepsilon$, 
where $K$ is the unit of energy, 
to specify systems.
Note that 
all quantities shown in figures such as $\fsp$ also have no dimensions.

\begin{figure}
\includegraphics[width=\columnwidth]{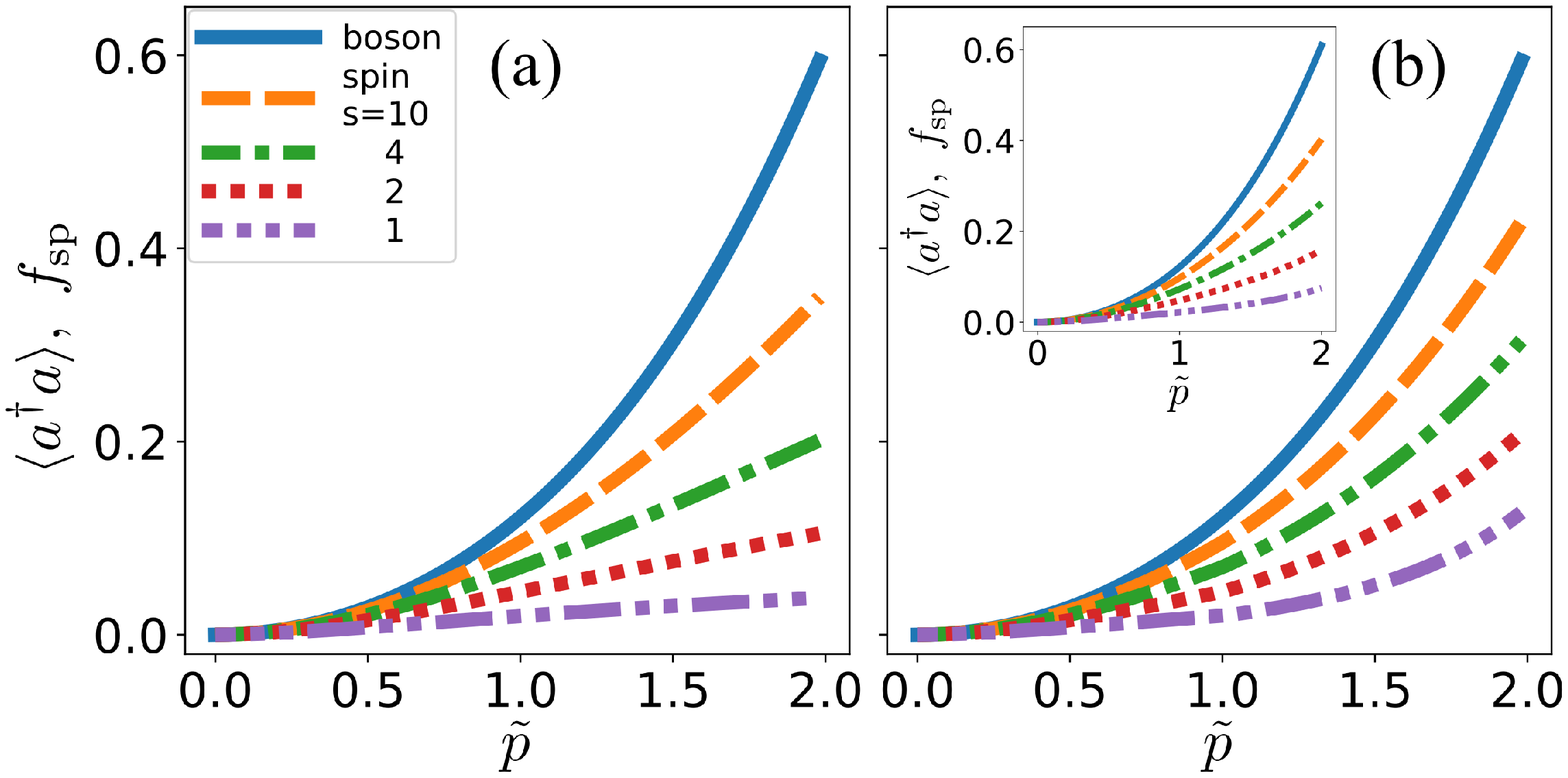}
\caption{\label{fig:photon_e=0}
			The photon number $\ave{\adag \ahat}$ of the ground state
			of a KPO as a function of $\tp$
			and the corresponding function $\fsp = s -\ave{\sx}$ 
			of the spin models
			for (a) $\alpha = 0$, (b) $\alpha = \alpha_c$, 
			and [inset of (b)] $\alpha^2$ equal to the exact $\ave{\adag \ahat}$.
			Both the two panels display the same photon number (solid).
			The curves for the spin models are  
			for $s=1$ (dashed double dotted), 2 (dotted), 4 (dashed dotted), 
			and 10 (dashed).
			All the systems are for $\tDelta = 1$ and $\tepsilon=0$.
			}
\end{figure}

Figure~\ref{fig:photon_e=0}~(a) shows 
the photon number $\ave{\adag \ahat}$ as a function of $\tp$
for $\tDelta = 1$ and $\tepsilon = 0$ 
and the corresponding function 
\begin{equation}
\fsp = s - \ave{\sx}
\end{equation}
of the spin models 
for $\alpha = 0$ 
and $s = 1$, 2, 4, and 10.
$\fsp$ is close to $\ave{\adag \ahat}$ at small $\tp$.
The difference appears and grows with $\tp$, 
but $\fsp$ captures the trend of $\ave{\adag \ahat}$.
$\fsp$ for larger $s$ shows smaller deviations from $\ave{\adag \ahat}$.
$\fsp$ for $\alpha = \alpha_c$ and equal to the exact $\sqrt{\ave{\adag \ahat}}$ 
are shown in Fig.~\ref{fig:photon_e=0}~(b) and its inset.
The model with $\alpha = \alpha_c$ suppresses the deviation of $\fsp$
from $\ave{\adag \ahat}$ more than the one with $\alpha = 0$
and behaves like the one with $\alpha$ equal to the exact $\sqrt{\ave{\adag \ahat}}$.

\begin{figure}
\includegraphics[width=\columnwidth]{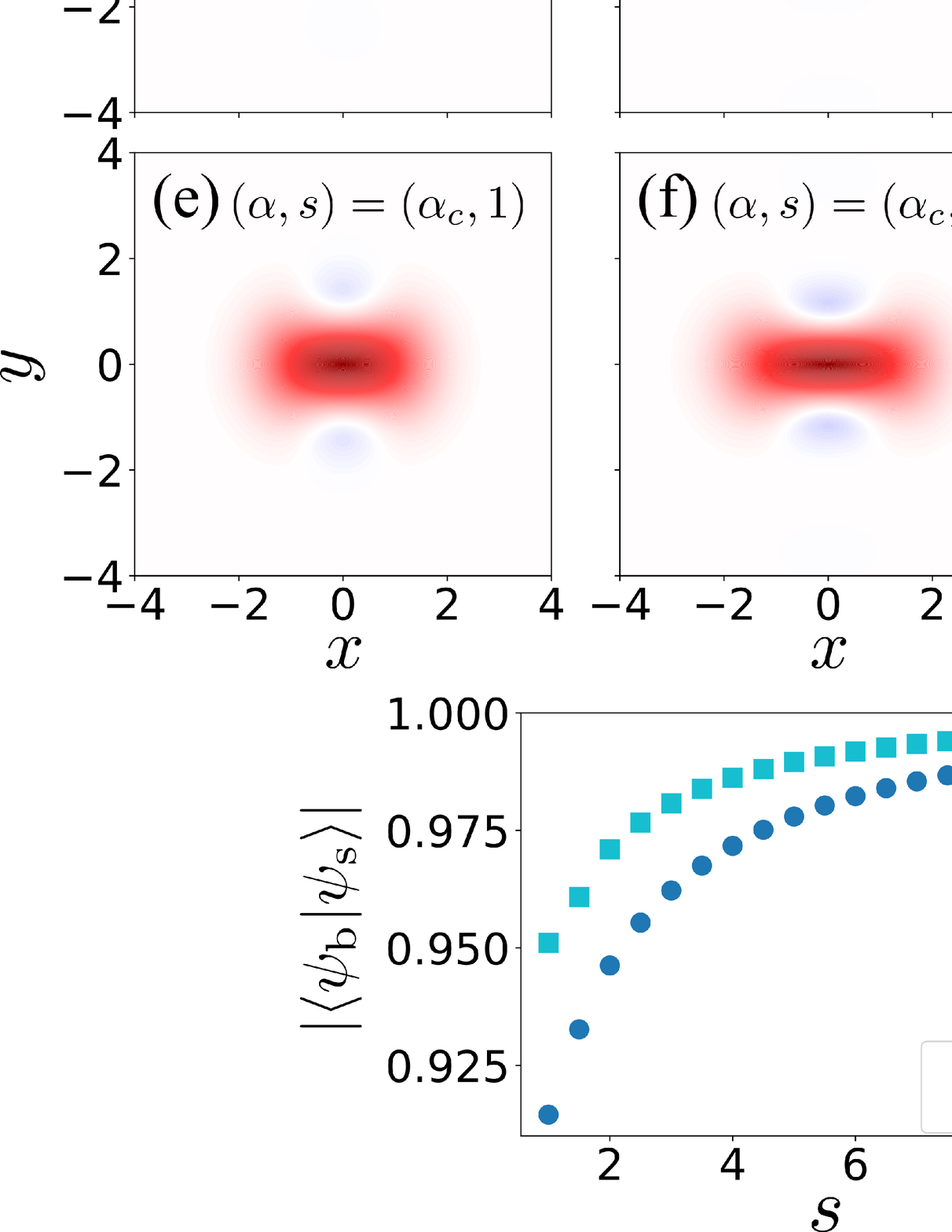}
\caption{\label{fig:wigner_e=0}
			The Wigner function of the ground state 
			of (a) a KPO 
			and the corresponding function of the spin models
			for (b)~$(\alpha, s) = (0, 1)$, (c)~$(0, 4)$, (d)~$(0, 10)$, 
			(e)~$(\alpha_c, 1)$, (f)~$(\alpha_c, 4)$, and (g)~$(\alpha_c, 10)$.
			$x$ and $y$ denote 
			$\sqrt{2}\text{Re}(\alpha)$ and $\sqrt{2}\text{Im}(\alpha)$ 
			for a coherent state $\ket{\alpha}$, respectively.
			(h) The overlap $| \langle \psib | \psis \rangle |$ 
			of bosonic and spin ground states
			as a function of $s$
			for $\alpha = 0$ (circle) and $\alpha_c$ (square).
			All the systems are for $\tDelta = 1$, $\tp = 2$ and $\tepsilon=0$.
			}
\end{figure}

Figure~\ref{fig:wigner_e=0}~(a) shows the Wigner function~\cite{D.Walls2008} 
of the ground state of a KPO for $\tDelta = 1$, $\tp=2$, and $\tepsilon = 0$. 
The shape of the region for positive values shows a sign of bifurcation.
The function also has negative values 
that is the evidence of quantum superposition.
In order to calculate the corresponding function of the spin models, 
we replace the density matrix of the KPO 
with that of the spin models.
Negative values are not clearly found in the resulting function
for $\alpha=0$ and $s = 1$ [Fig.~\ref{fig:wigner_e=0}~(b)], 
while they are found for larger $s$ [Figs.~\ref{fig:wigner_e=0}~(c) and (d)].
The spin model with $\alpha = \alpha_c$ takes negative values 
for $s=1$ [Fig.~\ref{fig:wigner_e=0}~(e)].
As $s$ increases [Figs.~\ref{fig:wigner_e=0}~(f) and (g)], 
the positive-value region is squeezed, 
and the function becomes more similar to that of the bosonic model.
The function similarity is quantified 
by the state overlap $| \langle \psib | \psis \rangle |$, 
where $\ket{\psib}$ and $\ket{\psis}$ denote the ground states of 
bosonic and spin models, respectively.
We define the overlap by
\begin{equation}
| \langle \psib | \psis \rangle |
= \left| \sum_{n=0}^{2s} \langle \psib | n \rangle {}_x\langle s - n, s | \psis \rangle  \right|, 
\end{equation}
where $\adag \ahat \ket{n} = n\ket{n}$ 
and $\sx\ket{s-n, s}_x = (s-n)\ket{s-n, s}_x$.
Figure~\ref{fig:wigner_e=0}~(h) demonstrates that 
the overlap for $\alpha = \alpha_c$ increases with $s$ 
and is always larger than that for $\alpha = 0$.
Note that 
the Wigner function for $s = 1/2$ does not show squeezing 
and the negative-value region (not shown), 
since the nonlinear terms in the Hamiltonian 
become constants when $s=1/2$.

\begin{figure}
\includegraphics[width=\columnwidth]{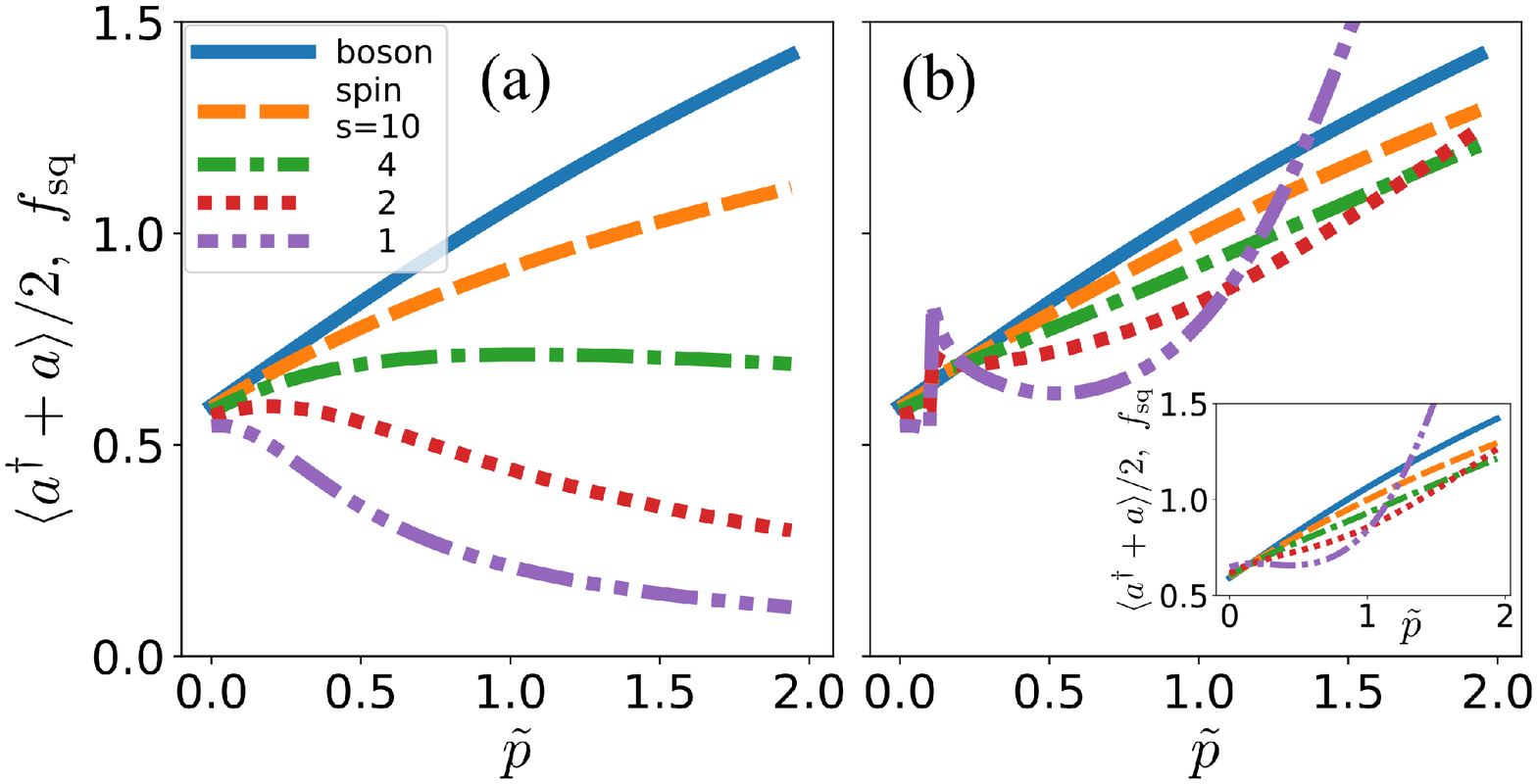}
\caption{\label{fig:quad_e=0.1}
			The quadrature amplitude $\ave{\adag + \ahat}/2$ of the ground state 
			of a KPO as a function of $\tp$
			and the corresponding function $\fsq =(2s - \alpha^2)^{-1/2} \ave{\sz}$ 
			of the spin model [Eq.~(\ref{eq:H_spin})]
			for (a) $\alpha = 0$, (b) $\alpha = \alpha_c$, 
			and [inset of (b)] $\alpha^2$ equal to the exact $\ave{\adag \ahat}$.
			Both the two panels display the same quadrature amplitude (solid). 
			The curves for the spin models are  
			for $s=1$ (dashed double dotted), 2 (dotted), 4 (dashed dotted), 
			and 10 (dashed).
			All the systems are for $\tDelta = 0$ and $\tepsilon=0.1$.
			}
\end{figure}

We also investigate the systems with coherent driving $\epsilon$, 
where the quadrature amplitude $\ave{\adag + \ahat}/2$ can have a finite value.
Figure~\ref{fig:quad_e=0.1}~(a) shows $\ave{\adag + \ahat}/2$ 
as a function of $\tp$
for $\tDelta = 0$ and $\tepsilon = 0.1$ 
and the corresponding function 
\begin{equation}
\fsq = \frac{\ave{\sz}}{\sqrt{2s - \alpha^2}}
\end{equation}
of the spin models 
for $\alpha = 0$ and $s = 1$, 2, 4, and 10.
The curves of $\fsq$ largely deviate
from the curve of $\ave{\adag +\ahat}/2$ except for $\tp \simeq 0$.
We find in Fig.~\ref{fig:quad_e=0.1}~(b) that 
$\fsq$ for $\alpha = \alpha_c$
and $s \ge 2$ do not have such a large difference even at large $p$, 
but that for $s=1$ has a steep peak at $\tp \simeq 0.1$ 
caused by the discontinuous change in $\alpha_c$ 
at $\tp = \tDelta + \tepsilon$, 
which is more accentuated in $1 - \alpha^2/2s$ for smaller $s$.
The similar curves, 
but without the steep peak, 
are obtained from the models 
with $\alpha$ equal to the exact $\sqrt{\ave{\adag \ahat}}$
[inset of Fig.~\ref{fig:quad_e=0.1}~(b)].
The result demonstrates that
the transformation based on the expansion 
with $\alpha_c$ gives pretty good spin models 
that can describe physical quantities of a KPO 
at least qualitatively.

\begin{figure}
\includegraphics[width=\columnwidth]{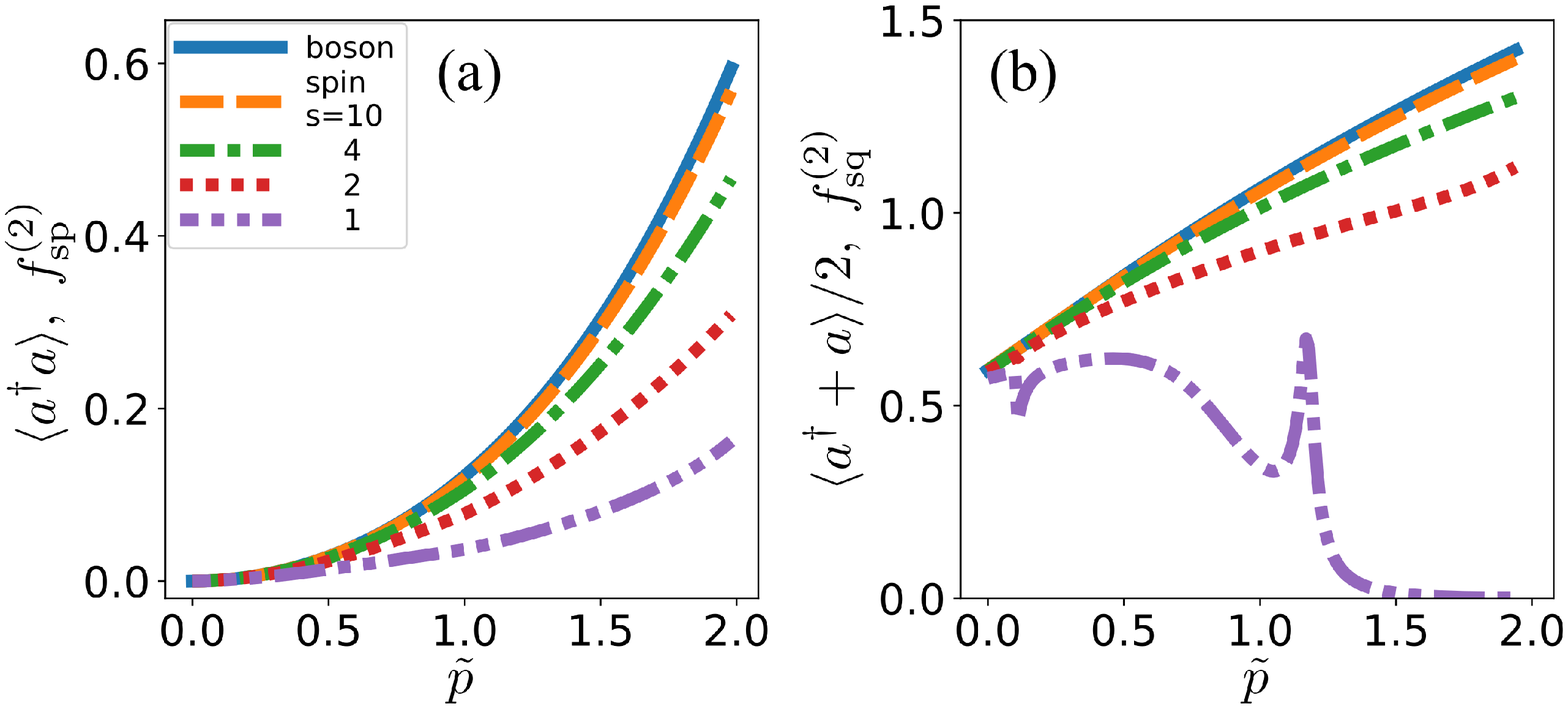}
\caption{\label{fig:second}
			(a) The photon number $\ave{\adag \ahat}$ 
			and (b) quadrature amplitude $\ave{\adag + \ahat}/2$ 
			of the ground state of a KPO as a function of $\tp$
			and the corresponding functions 
			(a) $\fsp^{(2)} = s -\ave{\sx}$ 
			and (b) $\fsq^{(2)}$ [Eq.~(\ref{eq:(adag+ahat)/2_2})]
			of the spin models including the second order terms 
			[Eq.~(\ref{eq:H_spin_second})] 
			for $\alpha = \alpha_c$.
			The solid curves show the results of the KPO.
			The curves for the spin models are  
			for $s=1$ (dashed double dotted), 2 (dotted), 4 (dashed dotted), 
			and 10 (dashed).
			$\tDelta$ and $\tepsilon$ are set to the same as those 
			for (a) Fig.~\ref{fig:photon_e=0} and (b) Fig.~\ref{fig:quad_e=0.1}, 
			i.e., (a) $(\tDelta, \tepsilon) = (1, 0)$ and (b) (0, 0.1), respectively.
			}
\end{figure}

Let us look at the spin models 
including the second order terms for $\alpha = \alpha_c$ 
that is given in Appendix.
We represent the spin counterparts 
of photon number and quadrature amplitude 
for this case
as $\fsp^{(2)}$ and $\fsq^{(2)}$, respectively.
The expression of $\fsq^{(2)}$ is different from $\fsq$ 
and is given by the right hand side of Eq.~(\ref{eq:(adag+ahat)/2_2}) in Appendix.
As shown in Fig.~\ref{fig:second}~(a), 
all the curves of $\fsp^{(2)}$ become closer to $\ave{\adag \ahat}$
compared to $\fsp$ [Fig.~\ref{fig:photon_e=0}~(b)].
$\fsq^{(2)}$ for $s = 4$ and 10 [Fig.~\ref{fig:second}~(b)]
also become closer to $\ave{\adag + \ahat}/2$ 
compared to $\fsq$ [Fig.~\ref{fig:quad_e=0.1}~(b)], 
but for $s=1$ and 2 show no such improvement.
This finding demonstrates that 
the transforamtion including higher order terms can give more accurate description, 
but the accuracy is not necessarilly improved for small $s$.

\subsection{Two interacting KPOs}

Let us move to interacting KPOs.
The simplest case is two KPOs 
governed by Eq.~(\ref{eq:H_KPOs}) for $N = 2$, $J_{12} = J_{21} = J$, and $\xi_0 =1$, 
\begin{equation}
\Hbtwo 
= \hat{H}_{\text{b},1} + \hat{H}_{\text{b}, 2} -J (\adag_1 \ahat_2 + \ahat_1 \adag_2),
\label{eq:H_2KPOs}
\end{equation}
where $\Hbi$ for $i = 1$ and 2 are given by Eq.~(\ref{eq:H_KPO_i}).
This system is transformed to
\begin{equation}
\Hstwo
= \hat{H}_{\text{s}, 1} + \hat{H}_{\text{s}, 2}
-\frac{2J}{2s - \alpha^2} 
\left( \sz_1 \sz_2 + \sy_1 \sy_2 \right), 
\label{eq:H_2spins}
\end{equation}
where $\Hsi$ for $i = 1$ and 2 is given by Eq.~(\ref{eq:H_spin}).
This spin model is based on the first order terms in the expansion.
As mentioned in Sec.~\ref{sec:transform}, 
$\sz_1 \sz_2$ might be dominant in the coupling.
We also investigate the ground state of
\begin{equation}
\Hstwop
= \hat{H}_{\text{s}, 1} + \hat{H}_{\text{s}, 2}
-\frac{2J}{2s - \alpha^2} 
\sz_1 \sz_2
\label{eq:H_2spins_zz}
\end{equation}
to examine the effects of ignoring $\sy_1 \sy_2$.
We set $\alpha$ to $\alpha_c$ for $\tepsilon=0$, namely
\begin{equation}
\alpha_{c0} =
\sqrt{ \tp-\tDelta} \theta(\tp-\tDelta), 
\label{eq:alpha_c0}
\end{equation}
where $J$ is not included as well as $\tepsilon$.
To incorporate the effects of coherent driving and interaction
into $\alpha$, 
we need the sign of quadrature amplitude of the ground state, 
which is what we aim to obtain with QA.
Therefore incorporating those effects is not fit 
for the purpose of constructing the model, 
and we adopt $\alpha_{c0}$ in Eq.~(\ref{eq:alpha_c0}) 
for $\alpha$.
When we investigated a single KPO in Sec.~\ref{sec:1KPO}, 
we included $\epsilon$ in $\alpha_c$, 
since the sign of quadrature amplitude in that case was trivial.

\begin{figure}
\includegraphics[width=\columnwidth]{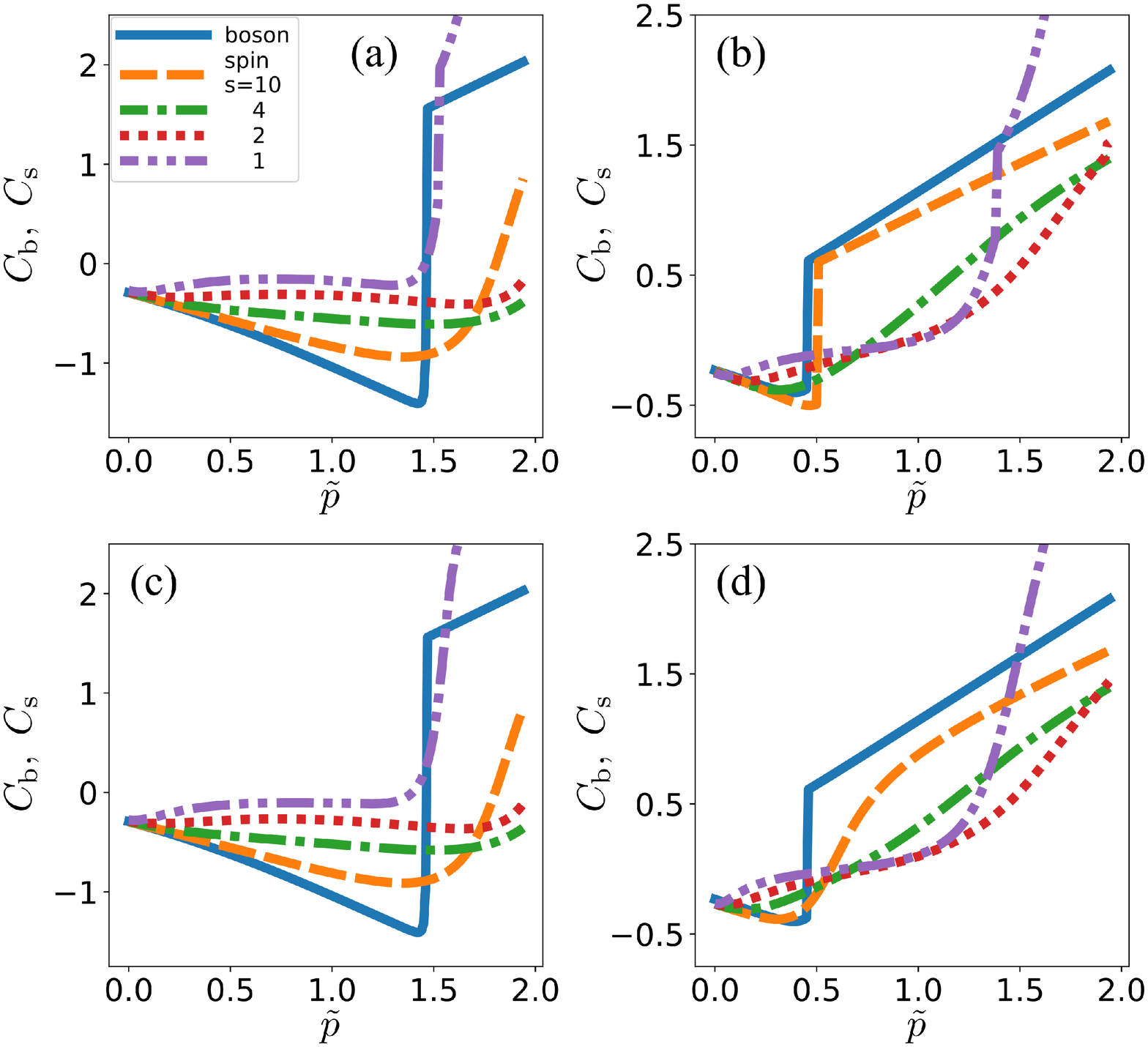}
\caption{\label{fig:correlate_2KPOs_frustration}
			The correlation 
			$\Cb = \ave{(\adag_1 + \ahat_1)(\adag_2 + \ahat_2)}/4$ 
			of quadrature amplitude of the ground state 
			of two KPOs as a function of $\tp$
			and the corresponding function  
			$\Cs = \ave{\sz_1 \sz_2}/(2s - \alpha_{c0}^2)$ 
			of the spin models 
			for $\alpha = \alpha_{c0}$.
			$\Cs$ shown in (a) and (b) is calculated for $\Hstwo$, 
			and in (c) and (d) is for $\Hstwop$.
			$(\tepsilon_1, \tepsilon_2, \tJ)$ is set to 
			(a) $(0.1,-0.1, 0.08)$, 
			(b) $(0.1,-0.1, 0.12)$, 
			(c) $(0.1,-0.1, 0.08)$, 
			and (d) $(0.1,-0.1, 0.12)$.
			The curves for the spin models are  
			for $s=1$ (dashed double dotted), 2 (dotted), 4 (dashed dotted), 
			and 10 (dashed).
			$\tDelta = 0$ in all the systems.
			}
\end{figure}

We focus on the correlation of quadrature amplitude of KPOs,
\begin{equation}
\Cb 
= \frac{1}{4} 
\left\langle \left( \adag_1 + \ahat_1 \right) \left( \adag_2 + \ahat_2 \right) \right\rangle, 
\end{equation}
and the corresponding function of the spin models
\begin{equation}
\Cs = \frac{\ave{\sz_1 \sz_2}}{ 2s - \alpha_{c0}^2}.
\end{equation} 
Figure \ref{fig:correlate_2KPOs_frustration} shows $\Cb$ and $\Cs$ 
for $\tepsilon_1= 0.1$, $\tepsilon_2 = -0.1$, and $\tJ = J/K = 0.08$ and 0.12.
$\Cb$ and $\Cs$ at small $\tp$ take negative values 
due to the difference in sign of $\tepsilon_1$ and $\tepsilon_2$
and shift to positive values at an intermediate $\tp$.
This shift is caused by the change of dominant term 
from the coherent driving to the coupling.
$\Cs$ for $\tJ = 0.08$ and $s = 10$ follows $\Cb$,  
although $\Cs$ is more gradual than $\Cb$ 
[Fig.~\ref{fig:correlate_2KPOs_frustration}~(a)].
$\Cs$ for $\tJ = 0.12$ and $s = 10$ reproduces the shift in $\Cb$
[Fig.~\ref{fig:correlate_2KPOs_frustration}~(b)], 
which occurs at smaller $\tp$ than for $\tJ = 0.08$.
$\Cs$ for $s = 1, 2$, and 4 agree with $\Cb$ at small $\tp$, 
but they depart from $\Cb$ as $\tp$ increases.
Note that 
it is just a coincidence that 
the rapid increase of $\Cs$ for $\tJ = 0.08$ and $s = 1$ overlaps $\Cb$.
We also compute $\Cs$ of $\Hstwop$,  
where $\sy_1 \sy_2$ is ignored.
We find no significant difference 
between $\Cs$ of $\Hstwo$ and $\Hstwop$ for $\tJ = 0.08$ 
[Fig.~\ref{fig:correlate_2KPOs_frustration}~(c)],
while $\Cs$ of $\Hstwop$ for $\tJ = 0.12$ deviates from that of $\Hstwo$ 
around $\tp$ 
where $\Cb$ shifts to positive values
[Fig.~\ref{fig:correlate_2KPOs_frustration}~(d)].
It is natural that 
differences appear in rapid changes of the function at small $\tp$, 
since $\sz_1 \sz_2$ is expected to be dominant only at large $\tp$.
Figure~\ref{fig:correlate_2KPOs_frustration} demonstrates that
$\Cs$ agrees with $\Cb$ at small $\tp$, 
while the gap between $\Cb$ and $\Cs$ grows with $\tp$, 
but $\Cs$ begins increasing and  exhibits the similar trend to $\Cb$ at large $\tp$.
The larger $s$ is, 
the larger range of $\tp$ over which 
our spin model can describe KPOs well.
In addition, 
$\Hstwop$
is a good approximation of $\Hstwo$, 
except for the rapid changes at small $\tp$.
Importantly, 
we can extract the correct sign of correlations
of the two KPOs, 
which corresponds to the Ising spin configuration in QA, 
from our spin models at large $\tp$.

\subsection{Mean-field model}

We investigate the mean-field models 
to effectively find the collective behavior.
The mean-field model of KPOs with a coupling constant $J$ is 
\begin{equation}
\Hbmf
= \Hb - J x \left(\adag + \ahat \right) - iJy \left(\adag - \ahat \right), 
\end{equation}
where $\Hb$ is given by Eq.~(\ref{eq:H_KPO}).
The coordination number is included in $J$.
$x$ and $y$ are the real and imaginary parts, 
respectively, 
of $a$ 
that is determined by the self-consistent equation,
\begin{equation}
a = \bra{\psib (a)} \ahat \ket{\psib (a)}
\end{equation} 
where $\ket{\psib (a)}$ is the ground state of $\Hbmf$ for $a$.
When $\tp = 0$, 
we can find the ground states 
that break the symmetry in $\Hbmf$ for $\tp = 0$ 
involving a transformation $\ahat \to \ahat e^{i\phi}$ 
with an arbitrary real $\phi$.
When analyzing such states, 
it is sufficient to consider only the ground state 
with $x \ge 0$ and $y = 0$.
When $\tp > 0$,
the pump suppresses $y$, 
and thus $y = 0$. 
We therefore simplify the Hamiltoinan as
\begin{equation}
\Hbmf
= \Hb - J x \left(\adag + \ahat \right).
\label{eq:H_KPO_mf}
\end{equation}
The corresponding mean-field spin model is
\begin{equation}
\Hsmf
= \Hs 
- \frac{2J}{ 2s - \alpha^2} m^z \sz,
\label{eq:H_spin_mf}
\end{equation}
where $\Hs$ is given by Eq.~(\ref{eq:H_spin}), 
and $m^z$ is determined by
\begin{equation}
m^z 
= \bra{\psis (m^z)} \sz \ket{\psis (m^z)} 
\end{equation}
for the ground state $\ket{\psis (m^z)}$ 
of $\Hsmf$ for $m^z$.
We set $\alpha = \alpha_{c0}$ in Eq.~(\ref{eq:alpha_c0}).
When we solve the self-consistent equations recursively, 
the initial values of $x$ and $m^z$ are set to tiny positive ones 
to focus on the solutions for $x \ge 0$ and $m^z \ge 0$.
We do not discuss the validity of the mean-field approximation 
to describe many interacting KPOs.
That is beyond the scope of this paper.

\begin{figure}
\includegraphics[width=0.7\columnwidth]{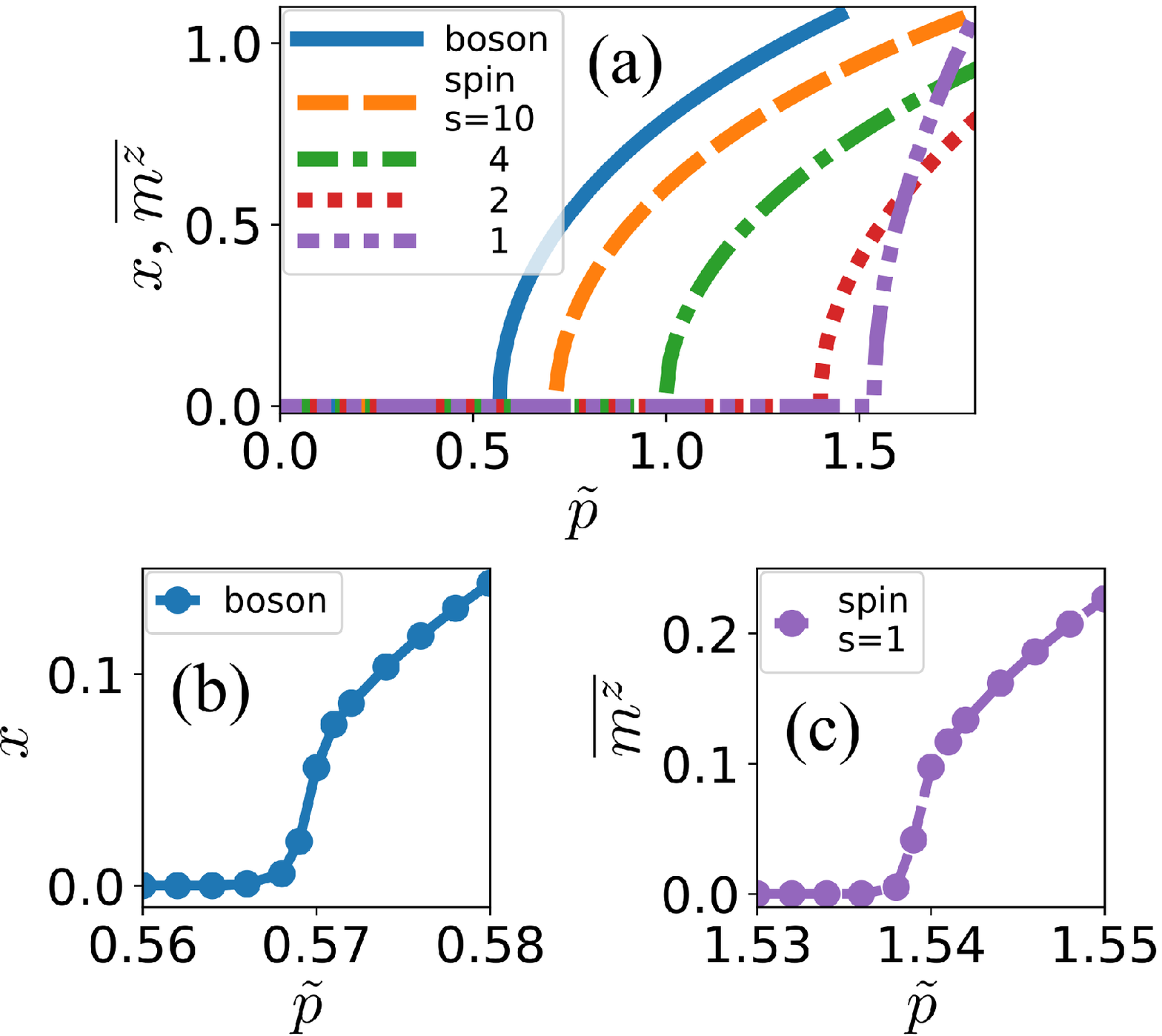}
\caption{\label{fig:phase_transition}
			(a) The expectation value $x$ of quadrature amplitude 
			of the ground state of the mean-field model for KPOs
			as a function of $\tp$ (solid)
			and the corresponding function $\overline{m^z}$ 
			of the mean-field spin models 
			for $\alpha = \alpha_{c0}$
			and $s=1$ (dashed double dotted), 2 (dotted), 4 (dashed dotted), 
			and 10 (dashed).
			The closeups of the onset of (b) $x$ 
			and (c) $\overline{m^z}$ for $s=1$ are also shown.
			All the systems are for $\tDelta = 0.4$, $\tepsilon = 0$, and $\tJ = 0.2$.
			}
\end{figure}

Both the expectation value $x$ of quadrature amplitde
and its spin counterpart 
$\overline{m^z} = (2s - \alpha^2)^{-1/2} m^z$ 
for small $\tDelta$
rise continuously from 0 as $\tp$ increases 
[Figs.~\ref{fig:phase_transition}~(a)--(c)].
The onset of $x$ or $\overline{m^z}$ can be interpreted as 
the continuous phase transition 
from the paramagnetic phase to the ferromagnetic phase.
The bifurcation induced by increasing $\tp$ 
underlies this phase transition.
The spin systems have similar magnetization curves 
to that of the bosonic system, 
although the critical points are shifted. 
\begin{figure}
\includegraphics[width=0.7\columnwidth]{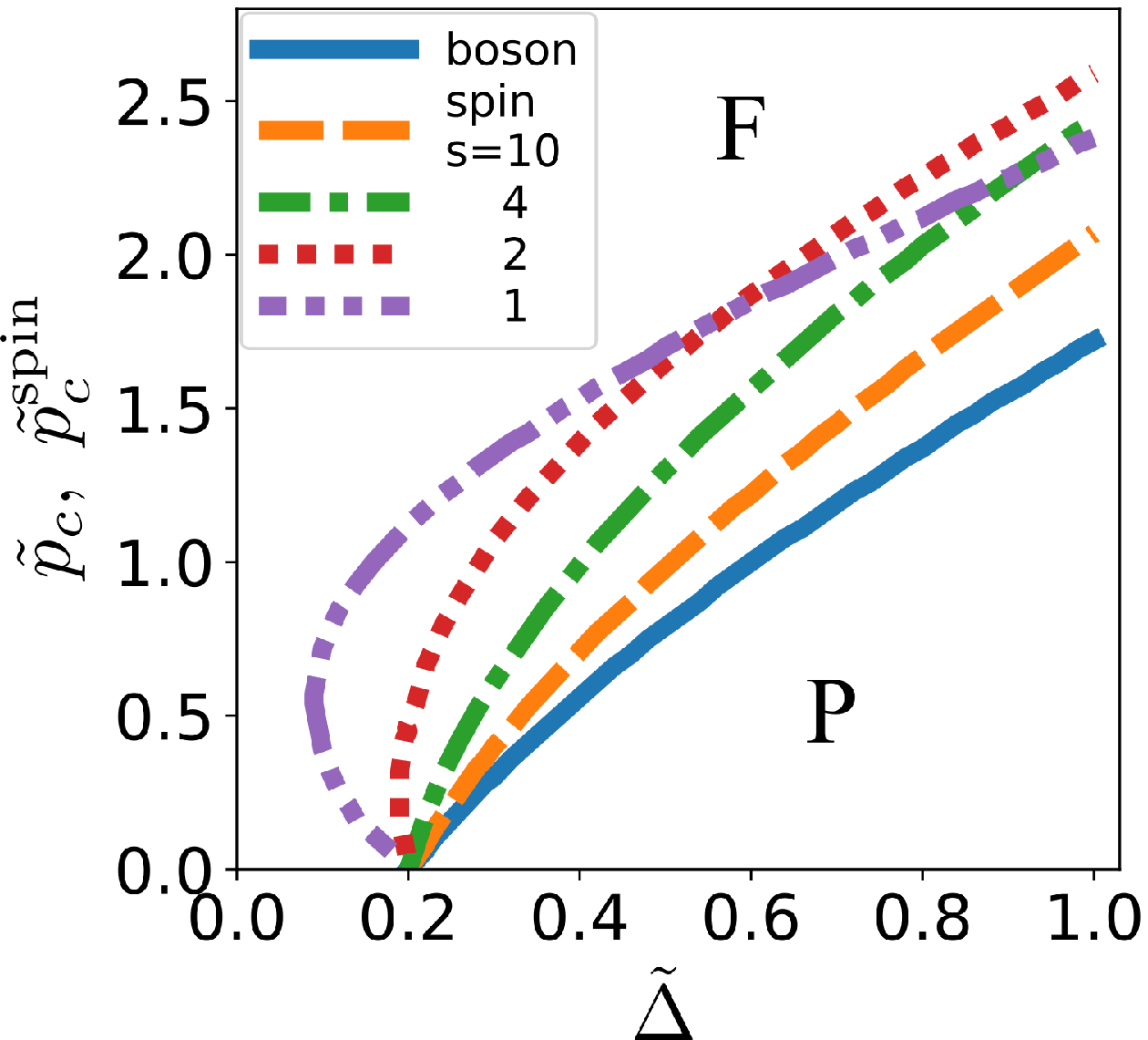}
\caption{\label{fig:phase_diagram}
			The critical pump amplitude $\tp_c$ 
			of the ground state of the mean-field model for KPOs
			as a function of $\tDelta$ (solid) 
			and the corresponding function $\tp_c^{\text{spin}}$ 
			of the mean-field spin models 
			for $\alpha = \alpha_{c0}$
			and $s=1$ (dashed double dotted), 2 (dotted), 4 (dashed dotted), 
			and 10 (dashed).
			The curves correspond to the phase boundaries 
			between the paramagnetic (P) and ferromagnetic (F)
			phases.
			All the systems are for $\tepsilon =0$ and $\tJ = 0.2$.
			}
\end{figure}
The critical points $\tp_c$ and $\tp_c^{\text{spin}}$  
of the bosonic and spin systems 
as a function of $\tDelta$ for $\tJ = 0.2$ are shwon 
in Fig.~\ref{fig:phase_diagram}.
The bosonic system for smaller $\tDelta$ than 0.2, say $\tDelta_c$, 
has a finite $x$ even at $\tp=0$, 
whereas that for $\tDelta > \tDelta_c$ lies in the paramagnetic phase at $\tp = 0$ 
and undergoes the phase transition at a finite $\tp_c$.
The spin systems for $s = 4$ and 10 
show similar phase boundaries, 
although the gap 
between $\tp_c^{\text{spin}}$ and $\tp_c$ grows 
as $\tDelta$ increases.
We also find that 
the spin systems for $s=1$ and 2 at $\tDelta \lesssim \tDelta_c$ 
exhibit the reentrant transition.
As $\tp$ is increased, 
the systems cross the phase boundary twice.
Their boundaries for larger $\tDelta$, however, show the same trend 
as that of the KPOs.

\section{Conclusion}
\label{sec:summary}

We have presented effective spin-$s$ models 
of Kerr-nonlinear parametric oscillators (KPOs) 
for quantum annealing (QA).
The detuning and coherent driving work as 
transverse and longitudinal fields in the spin systems, respectively.
The Kerr effect and the parametric driving 
turn to the nonlinear terms of $\sx$ and $\sz$, respectively. 
Although we truncate the expansion in the transformation, 
the spin models, in particular, for large $s$ at small $\tp$ 
show good correspondence 
to the bosonic model.
Even when $s$ is small but larger than 1/2, 
the spin models partially capture features of KPOs, 
e.g., bifurcation and phase transitions.
As the photon number of KPOs becomes large, 
the gap between the spin and bosonic models grows, 
but setting a parameter $\alpha$ to an approximate value of $\sqrt{\ave{\adag \ahat}}$ 
makes the gap smaller.
The gap can be also reduced 
by including higher order terms in the transformation.

The spin models presented in this paper are simple 
but qualitatively reproduce KPOs 
and clarify roles of parameters of KPOs in QA.
We expect that 
the models will help us understand the behavior of KPOs, 
in particular, 
when comparing QA with KPOs to 
conventional QA based on the transverse-field Ising model.
Improvements in quantitative accuracy would increase 
the usefulness of the method, 
but that is a topic for future work.

\appendix*
\section{Transformation with the second term in Eq.~(\ref{eq:sz})}
\label{sec:higher}

Equation (\ref{eq:sz}) reads
\begin{eqnarray}
\sz
= && \sqrt{2s -\alpha^2} \frac{\adag + \ahat}{2} 
\nonumber \\
&&- \frac{1}{4\sqrt{2s - \alpha^2}} 
\bigg[ \adag \ahat \frac{\adag + \ahat}{2} 
+ \frac{\adag + \ahat}{2} \adag \ahat 
\nonumber \\
&& -\left( 1 + \alpha^2 \right) \frac{\adag + \ahat}{2} \bigg] 
+ \cdots.
\label{eq:sz_second}
\end{eqnarray}
We assume that
$(\adag + \ahat)/2$ is represented as $\sz/\sqrt{2s-\alpha^2} + A_1$, 
where $\sz/\sqrt{2s-\alpha^2}$ is obtained from the transformation 
that only takes into account the first term in Eq.~(\ref{eq:sz_second}).
We substitute $\sz/\sqrt{2s-\alpha^2} + A_1$ 
into $(\adag + \ahat)/2$ in Eq.~(\ref{eq:sz_second}) 
including up to the second term
and ignore terms such as $A_1/ \sqrt{2s-\alpha^2}$
to obtain 
\begin{equation}
A_1 
= \frac{(2s - 1 - 2\alpha^2)\sz - \sx \sz - \sz \sx}{4(2s-\alpha^2)^{3/2}}
\end{equation}
and
\begin{eqnarray}
\frac{\adag + \ahat}{2}
= && \left[ \frac{5}{4\sqrt{2s - \alpha^2}} 
- \frac{1 + \alpha^2}{4(2s-\alpha^2)^{3/2}}\right] 
\sz 
\nonumber \\
&& - \frac{\sx \sz + \sz \sx}{4(2s-\alpha^2)^{3/2}}, 
\label{eq:(adag+ahat)/2_2}
\end{eqnarray}
where the higher order terms 
in powers of $1/(2s-\alpha^2)$ are dropped.
Equation (\ref{eq:sz}) also leads to 
\begin{eqnarray}
(\sz)^2
= && \frac{2s-\alpha^2}{2} 
\left( \frac{\adagtwo + \ahat^2}{2} + \adag \ahat + \frac{1}{2} \right) 
\nonumber \\
&& - \bigg[\frac{(\adag \ahat)^2}{2} 
+ \left( \adag \ahat \frac{\adagtwo + \ahat^2}{2} 
+ \frac{\adagtwo + \ahat^2}{2} \adag \ahat \right) 
\nonumber \\
&& - \frac{1 + 2\alpha^2}{4} \frac{\adagtwo + \ahat^2}{2} 
- \frac{\alpha^2}{2} \adag \ahat - \frac{\alpha^2}{4}
\bigg] + \cdots.
\label{eq:sz^2_second}
\end{eqnarray}
We substitute $- s + \sx - 1/2 + 2(\sz)^2/(2s- \alpha^2) + B_1$
into $(\adagtwo + \ahattwo)/2$ in Eq.~(\ref{eq:sz^2_second}), 
where $- s + \sx - 1/2 + 2(\sz)^2/(2s- \alpha^2)$ is obtained 
from the transformation 
that only takes into account the first line in Eq.~(\ref{eq:sz^2_second}).
We also ignore terms such as $B_1/ (2s-\alpha^2)$ 
to obtain
\begin{eqnarray}
B_1 
=&& \frac{( 2s - 1 - 2\alpha^2 ) (\sz)^2 - \sx (\sz)^2 - (\sz)^2 \sx}{(2s-\alpha^2)^2}
\nonumber \\
&&+ \frac{1}{4(2s-\alpha^2)}
\end{eqnarray}
and
\begin{eqnarray}
\frac{\adagtwo + \ahattwo}{2}
=&& - s + \sx - \frac{1}{2} 
+ \left[ \frac{3}{2s- \alpha^2} - \frac{1+\alpha^2}{(2s-\alpha^2)^2} \right] (\sz)^2 
\nonumber \\
&& + \frac{1}{4(2s-\alpha^2)}
- \frac{\sx (\sz)^2 + (\sz)^2 \sx}{(2s-\alpha^2)^2}, 
\label{eq:(adagtwo+ahattwo)/2_2}
\end{eqnarray}
where the higher order terms in powers of $1/(2s-\alpha^2)$ are dropped.
Equations~(\ref{eq:adagahat}), (\ref{eq:adagtwoahattwo}) 
(\ref{eq:(adag+ahat)/2_2}), and (\ref{eq:(adagtwo+ahattwo)/2_2})
are used to transform the Hamiltonian for a KPO [Eq.~(\ref{eq:H_KPO})] to 
\begin{eqnarray}
\hat{H}_{\text{spin}} 
=&& 
- \left[ \frac{3}{2s- \alpha^2} - \frac{1+\alpha^2}{(2s-\alpha^2)^2} \right] p (\sz)^2
\nonumber \\
&&- \left[ \Delta + p + \left(s - \frac{1}{2} \right) K \right] \sx 
+ \frac{K}{2} (\sx)^2
\nonumber \\
&&
-\left[ \frac{5}{2\sqrt{2s - \alpha^2}} 
- \frac{1 + \alpha^2}{2(2s-\alpha^2)^{3/2}}\right] \epsilon \sz
\nonumber \\
&&
+ \frac{\epsilon}{2(2s-\alpha^2)^{3/2}} \left( \sx \sz + \sz \sx \right)
\nonumber \\
&&
+\frac{p}{(2s-\alpha^2)^2} \left[ \sx (\sz)^2 + (\sz)^2 \sx \right].
\label{eq:H_spin_second}
\end{eqnarray}

\begin{acknowledgments}
The author thanks Y. Susa for helpful comments on the manuscript.
This paper is based on results obtained from a project, 
JPNP16007, 
commissioned by 
the New Energy and Industrial Technology Development Organization (NEDO), Japan.
We have used QuTiP~\cite{J.Johansson2012, J.Johansson2013}
in some of the numerical calculations.
\end{acknowledgments}

%

\end{document}